\documentclass[12pt]{article}

\parskip=\medskipamount            
\def\7#1#2{\mathop{\null#2}\limits^{#1}}        

\def\beee{\begin{equation}}
\def\eeee{\end{equation}}

\begin{document}
\bibliographystyle{unsrt}

\begin{center}
\textbf{Remarks on a challenge to the relation\\ between $CPT$ and Lorentz violation}\\
\vspace{5mm}
O.W. Greenberg\footnote{email address, owgreen@umd.edu}\\
\textit{Center for Fundamental Physics\\
Department of Physics \\
University of Maryland\\
College Park, MD~~20742-4111, US}\\
and \textit{Rockefeller University\\
123 York Avenue\\
New York, NY 10065}

University of Maryland Preprint PP-11-001\\
\end{center}

\begin{abstract} 
The objection to my theorem that violation of $CPT$ symmetry implies violation of Lorentz covariance
is based on a nonlocal model in which time-ordered products are not well defined. I used covariance of time-ordered
products as the condition for Lorentz covariance; therefore the proposed objection is not relevant to my result.
\end{abstract}

\section{Introduction}
In demonstrating that violation of $CPT$ symmetry implies violation of 
Lorentz covariance,~\cite{gre1} 
I explicitly assumed the properties of relativistic quantum field theory that are the basis of the Wightman 
formalism and Jost's theorem for the necessary and sufficient conditions for $CPT$ symmetry. (See reference
[3] of~\cite{gre1}.) I chose covariance of
of $T$ (or $r$ or $a$) products as the criterion for Lorentz covariance.  I also
implicitly assumed (i) that the $S$ matrix is well defined in the theory, 
which means that the in and out fields are related by a unitary $S$ matrix, 
$\phi^{out}(x) =S^{-1}\phi^{in}(x)S$, (ii) that the
theory has a finite number of fields,  and (iii) that the theory is formulated on ordinary (commutative) spacetime. 

A recent paper that challenges my result~\cite{cha} concerns a nonlocal model in which the $T$ products are 
not well-defined and thus is not a counter example to my result. In addition, the paper by 
R. Marnelius~\cite{mar} cited by the authors
for ``general
considerations on the causality and unitarity properties of nonlocal relativistic quantum field theories.''
states in the abstract  \textit{``This implies that the field equations do not yield unique
quantum solutions and in particular that the solutions with canonical incoming free fields are different
from the solutions with canonical outgoing free fields, and none of these solutions render the total
action stationary. \textbf{No meaningful S matrix can therefore be defined.}} (Boldface added by me.) 
\textit{It is also shown that this deficiency
cannot be corrected either by restricting the form function or by adding correction terms to the 
perturbation expansions.''} 

Further, the authors of~\cite{cha} cite two papers,~\cite{bar1} 
and~\cite{bar2}, that I criticized,~\cite{gre1}, 
as though I were the author of those papers. 

The challenge to my theorem does not question 
my derivation of the theorem based on the fact that violation of one
of the necessary and sufficient conditions for $CPT$ symmetry also violates one of the conditions for Lorentz covariance.

It is well known that 
local commutativity (or anticommutativity) of fields
implies covariance of time-ordered products. I also proved the converse of this result, that covariance
of time-ordered products implies local commutativity (or anticommutativity) of fields.~\cite{gre2} Since
I assumed covariance of the $T$ products, the fact that, as the authors state, local commutativity of fields
is violated in their model is a second way in which their nonlocal model 
violates the conditions of my theorem and thus is not relevant to my result.

\section{Discussion of other comments of~\cite{cha}}

Section 2 of the authors' paper refers to the model of~\cite{bar1} and~\cite{bar2}.  
These papers are their reference [15].
They accept my 
criticisms~\cite{gre1} of these two papers, but nontheless then refer to this model three times as 
though I had proposed it. Where they
should cite their reference [15], they cite my paper which is their reference [17]. This repeated interchanging of their
references [15] and [17] is highly misleading. They call the
model of~\cite{bar1} and~\cite{bar2} ``utmost pathological'' and repeat further criticisms of the model 
that I had already made~\cite{gre1}.
They correctly state that because observables in this model do not 
commute at space-like separation, the
proof of the spin-statistics relation is not valid. They then state ``there is \textit{no concept of
spin} to start with altogether.'' This is not correct; one can define spin by the transformation property
of the field under the Lorentz group or under the rotation group with support on only one mass shell 
in the same way one can define spin for a field with support on both mass shells.

At the end of their section 4 the authors assert ``With such a $CPT$-violating interaction ... the quantum corrections
due to the combined interactions could lead to different properties for the particle and antiparticle, including their
masses.'' The authors do not provide a calculation to support this assertion, nor do they explain how their model
would evade the specific detailed problems, including violations of Lorentz covariance, pointed out in my paper~\cite{gre1}.

\section{Summary}
The proposed counter-example of~\cite{cha} is not relevant to my theorem because I assumed Lorentz covariance 
of $T$ products as a condition of my theorem and the $T$ products in their model are not covariant.
In addition, according to a reference~\cite{mar} cited by the authors, their model does not have a properly defined
S matrix at all. 

\section{Acknowledgements}
It is a pleasure to thank Professor Nicola N. Khuri for his hospitality at the Rockefeller University
where some of this work was carried out. I thank Aiyalam Balachandran, Thomas Cohen, Steven Cowen, Alan Kostelecky,
Nicola N. Khuri, Raman Sundrum and Daniel Zwanziger
for helpful discussions. This work was 
supported in part by a sabbatical leave granted by the Department of Physics, University of Maryland.


\begin{thebibliography}{99}

\bibitem{gre1} O.W. Greenberg, Phys. Rev. Lett. 89, 231602 (2002).

\bibitem{cha} M. Chaichian, A.D. Dolgov, V.A. Novikov and A. Tureanu, Phys. Lett. B 699, 177 (2011); arXiv:11030168.

\bibitem{mar} R. Marnelius, Phys. Rev. D 10 3411, (1974). 

\bibitem{gre2} O.W. Greenberg, Phys. Rev. D 73, 087701 (2006).

\bibitem{bar1} G. Barenboim, L. Borissov, J.D. Lykken and A.Y. Smirnov, JHEP 0210, 001, (2002).

\bibitem{bar2} G. Barenboim, L. Borissov and J.D. Lykken, Phys. Lett B 106 (2002).

\end{thebibliography}
\end{document}